\documentclass[fleqn,10pt]{wlscirep}

\title{Coherency-broken Bragg filters: surpassing on-chip rejection limitations}

\author[1]{Dorian Oser}
\author[2]{Florent Mazeas}
\author[1]{Xavier Le Roux}
\author[1,$\bigtriangleup$]{Diego P\'{e}rez-Galacho}
\author[2]{Olivier Alibart}
\author[2]{Sebastien Tanzilli}
\author[2]{Laurent Labont\'{e}}
\author[1]{Delphine Marris-Morini}
\author[1]{Laurent Vivien}
\author[1]{\'{E}ric Cassan}
\author[1,*]{Carlos Alonso-Ramos}

\affil[1]{Centre for Nanoscience and Nanotechnology, CNRS, Universit\'{e} Paris-Sud, Universit\'{e} Paris-Saclay, C2N – Orsay, 91405 Orsay Cedex, France}
\affil[2]{Universit\'{e} C\^{o}te d’Azur, CNRS, Institut de Physique de Nice - INPHYNI, Parc Valrose, 06108 Nice Cedex 2, France}

\affil[*]{carlos.ramos@u-psud.fr}

\begin{abstract}
Selective on-chip optical filters with high rejection levels are key components for a wide range of advanced photonic circuits. However, maximum achievable rejection in state-of-the-art on-chip devices is seriously limited by phase errors arising from fabrication imperfections. Due to coherent interactions, unwanted phase-shifts result in detrimental destructive interferences that distort the filter response, whatever the chosen strategy (resonators, interferometers, Bragg filters, etc.). Here we propose and experimentally demonstrate a radically different approach to overcome this fundamental limitation, based on coherency-broken Bragg filters. We exploit non-coherent interaction among modal-engineered waveguide Bragg gratings separated by single-mode waveguides to yield effective cascading, even in the presence of fabrication errors. This technologically independent approach allows seamless combination of filter stages with moderate performance, providing a dramatic increase of on-chip rejection. Based on this concept, we experimentally demonstrate on-chip non-coherent cascading of Si Bragg filters with a record light rejection exceeding 80 dB in the C-band.
\end{abstract}
\begin{document}

\flushbottom
\maketitle
%
%
\thispagestyle{empty}



The silicon-on-insulator (SOI) platform allows realizing miniaturized optical circuits that can be fabricated in existing CMOS facilities. These high-performance photonic devices have a great potential for a plethora of applications, including datacom \cite{boeuf2016silicon}, sensing \cite{fernandez2016last, janz2013photonic}, and quantum information \cite{knill2001scheme,mazeas2016high}. While many highly efficient components have already been demonstrated in the silicon photonics platform, e.g. fiber-chip couplers \cite{cheben2015broadband}, fast modulators \cite{ziebell201240} and photodetectors \cite{vivien2012zero}, the realization of high rejection wavelength filters remains a challenge. Indeed, the lack of on-chip optical filters with strong rejection hinders the full integration of some advanced nonlinear circuits. One clear example is silicon photon-pair sources, with a great potential for applications in quantum key distribution \cite{jouguet2013experimental} and optical quantum computing \cite{knill2001scheme}. These sources exploit spontaneous four-wave mixing in Si micro-ring resonators to generate multi-spectral entangled photon-pairs from a strong optical pump \cite{azzini2012ultra, jiang2015silicon, grassani2015micrometer, mazeas2016high}. However, due to the substantially higher pump intensity compared to that of the photon-pairs, such Si-based sources require on-chip rejections exceeding 100\,dB \cite{piekarek2017high}. This stringent rejection requirement lies beyond the capabilities of current Si wavelength filters, precluding full integration of Si-based photon-pair sources. 

A myriad of optical filters has been reported for the silicon photonics technology, including Bragg grating filters \cite{wang2011uniform,perez2017optical,klitis2017high}, cascaded micro-resonators \cite{xia2007ultra,dong2010ghz} and Mach-Zehnder interferometers \cite{ding2011bandwidth,horst2013cascaded}. Although theoretical designs can achieve remarkably large rejection levels, most practical implementations are limited to the 30-60\,dB range \cite{xia2007ultra,dong2010ghz,ding2011bandwidth,wang2011uniform,horst2013cascaded,wang2015subwavelength,zou2016tunable,perez2017optical,klitis2017high}. The main limiting factor to the achievable on-chip optical rejection currently lies in fabrication imperfections. More specifically, the high index contrast of the SOI platform make these circuits very sensitive to fabrication errors, as small deviations in waveguide width and height strongly affect the propagation constant of light, resulting in large phase errors \cite{lu2017performance}. This detrimental effect distorts the filter response, thus compromising its rejection capability. It is worth mentioning that conventional multi-stage filters are also affected by this limitation, as phase errors actually shift the wavelength response of each filter section and produce detrimental destructive interferences, offsetting the benefits of the cascading. These drawbacks have been partially alleviated by implementing active phase-tuning in multi-stage filters. For instance 60\,dB rejection has been shown on a single chip with cascaded Mach-Zehnder interferometers \cite{piekarek2017high}, and 100\,dB has been demonstrated for a tenth order micro-ring-based filter \cite{ong2013ultra}. Still, this approach complicates device fabrication and operation, as it requires implementation of tuning circuitry and continuous monitoring of the filter response to maintain proper performance. 

\begin{figure}[htbp]
	\centering
	\includegraphics[width=12cm]{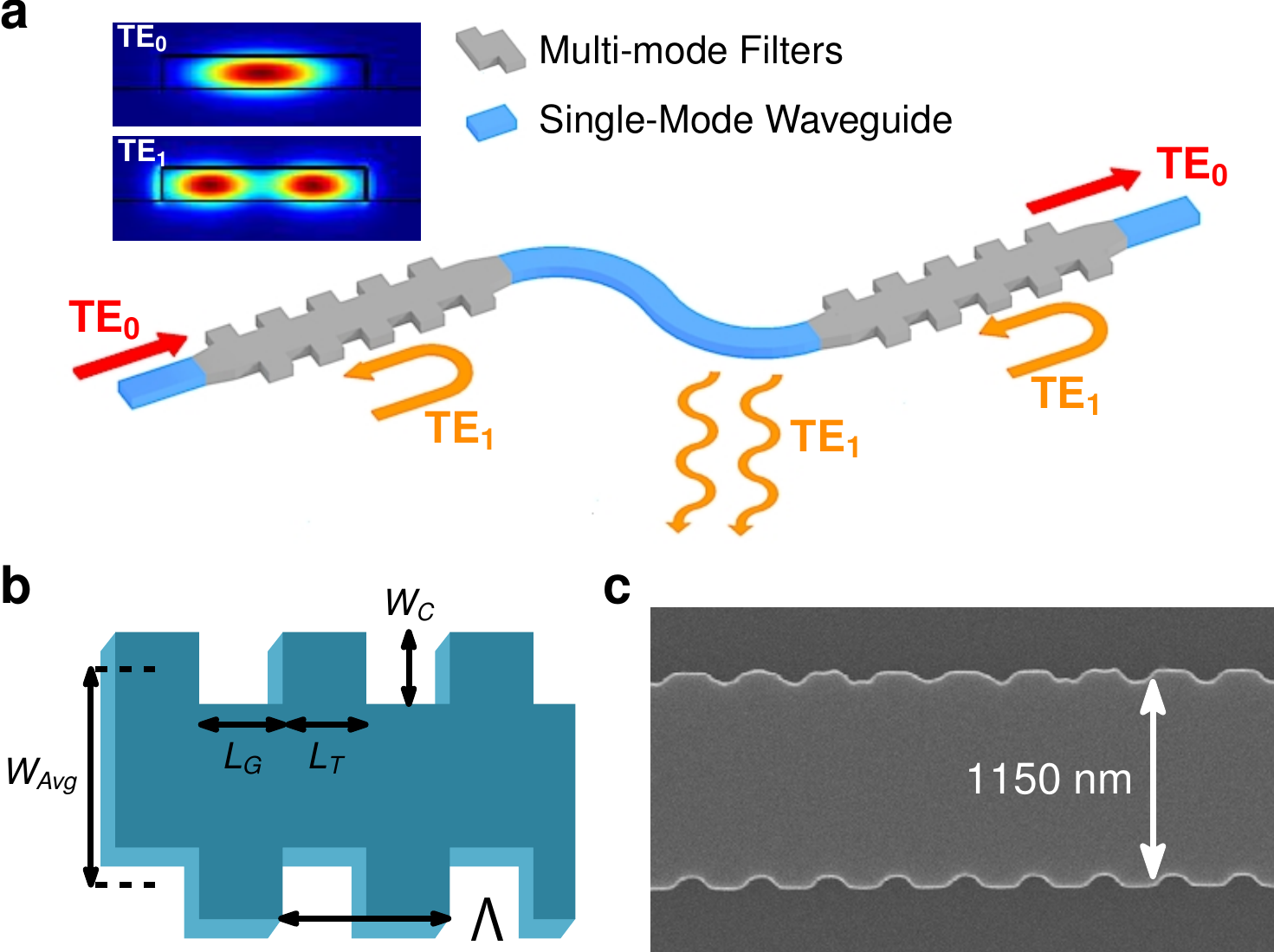}
	\caption{(a) Schematic view of proposed cascaded filter. Fundamental mode ($\mathrm{TE_0}$) is back-reflected into first order mode ($\mathrm{TE_1}$). Single-mode waveguide sections separating adjacent filters radiate the back-reflected $\mathrm{TE_1}$ mode away, precluding coherent interaction among different stages. (b) Schematic of shifted Bragg geometry providing Bragg back-reflections in higher order $\mathrm{TE_1}$ mode. (c) Scanning electron microscope (SEM) image of a fabricated shifted grating.}
	\label{fig:Schema}
\end{figure}

Here, we present a new strategy for the on-chip implementation of high-rejection multi-stage filters, free of active circuit control. The proposed approach, schematically depicted in Fig. \ref{fig:Schema}, exploits modal engineering in waveguide Bragg gratings to achieve non-coherent cascading, making the device immune to relative phase errors among the stages. It turns out that such strategy permits overcoming one of the major limitations of on-chip filters. The waveguide gratings are shaped to yield Bragg back-reflections propagating in a high-order spatial mode. These back-reflections are radiated away in single-mode waveguides interconnecting adjacent filter stages, precluding coherent interaction. This generic strategy allows the implementation of high-rejection filters by all-passive cascading of modal-engineered Bragg gratings with relaxed performance requirements. Hence, the broken-coherency Bragg filter approach proposed here opens a completely new route for the implementation of high-performance on-chip optical filters.

\section*{Results}

Waveguide Bragg gratings reflect light back into the input waveguide by constructive interference of partial reflections in each period \cite{liu2009photonic}. This resonant back-reflection occurs when the Bragg phase-matching condition, $ \lambda_o = \Lambda 2 n_{eff} / p $, is satisfied. Here, $\lambda_o$ is the Bragg resonance wavelength, $\Lambda$ the grating period, $n_{eff}$ the effective index of the mode propagating through the grating and $p$ is the Bragg order. On the other hand, from coupled mode theory, it is known that the rejection ($R$) and bandwidth ($\Delta \lambda$) of the filter are related to the grating geometry and filter length ($L_F$) by \cite{CMT}:
\begin{equation}
R = \tanh^2(\kappa L_F),
\label{eq:R}
\end{equation}
\begin{equation}
\Delta \lambda = \frac{\lambda_0^2}{\pi n_g} \sqrt{\kappa^2 + \frac{\pi^2}{L_F^2}}.
\label{eq:BW}
\end{equation}
Note that the group index ($n_g$) and the coupling coefficient ($ \kappa $) here are mainly governed by the grating geometry. Ideally, arbitrarily large rejections can be achieved just by implementing a sufficiently long filter. In practice, achievable rejection saturates beyond a certain filter length \cite{wang2011uniform,perez2017optical,klitis2017high}. To achieve the theoretical rejection level, partial reflections in all periods of the filter have to interfere constructively. However, even very small fabrication imperfections alter the phase along the filter, distorting the constructive interference. Such errors accumulate with increasing filter length, setting the saturation level. Cascading conventional Bragg filters does not address this issue, as back-reflections from each stage still need to propagate in phase through the previous gratings. This has two detrimental consequences: first, the effectiveness of cascading is affected by relative phase shifts between stages; second, fabrication imperfections in each section distort back-reflections generated by the following sections. To overcome this limitation, we propose a new multi-stage filter strategy combining multi-mode Bragg gratings and single-mode interconnection waveguides (see Fig. \ref{fig:Schema}). The Bragg gratings are designed to yield back-reflections propagating in the first higher order mode. Then, back-reflections are radiated away in the single-mode interconnection waveguide section, avoiding propagation through previous gratings. This way we preclude coherent interaction among filter stages, circumventing the detrimental effect of cumulative phase errors, thus achieving effective cascading.

To demonstrate this concept, we used SOI substrate with 220-nm-thick guiding Si layer. For the implementation of the multi-mode Bragg grating, we selected a fully-etched process and a shifted-teeth geometry, as presented in Fig. \ref{fig:Schema}(b). The grating lattice is defined by the average waveguide width ($W_{Avg}$), the corrugation depth ($W_{C}$), the length of the teeth ($L_{T}$) and gaps ($L_{G}$) and the period ($\Lambda=L_{T}+L_{G}$). By shifting the corrugation in one side of the grating half a period with respect to the other, this grating geometry precludes Bragg reflections in the fundamental mode \cite{Shifted}, while providing the asymmetry required to excite Bragg back-reflections in the first higher order mode \cite{qiu2016silicon}. 

The phase-matching condition for this hybrid fundamental-to-first-mode Bragg reflection can be expressed as $ \lambda_0 = \Lambda (n_{eff}^1 + n_{eff}^2) / 2 $. In this case $ n_{eff}^{1} $ is the effective index of the fundamental mode and $ n_{eff}^{2} $ is the effective index of the first higher-order mode. We designed the shifted Bragg grating to operate with transverse-electric (TE) polarized light around 1550 nm wavelength using modal analysis and finite-difference time domain (FDTD) tools from commercial software \cite{Lumerical}. The proposed design has a period $ \Lambda $ of 290 nm, an average waveguide width $ W_{Avg} $ of 1150 nm, a corrugation $ W_{C} $ of 50 nm and a duty cycle of 50\% ($L_{T}=L_{G}=145$\,nm).

The different sets of shifted Bragg filters were fabricated in SOI wafers comprising a $220$\,nm thick silicon and a $ 2\,\mu$m buried oxide layer. Air is used as top cladding to avoid propagation of the transverse-magnetic (TM) modes that may limit the measured rejection. Figure \ref{fig:Schema}(c) shows the scanning microscope image of one of the gratings. It can be noticed that the filter teeth are rounded, with local defects. We verified by simulation that rounding of the grating teeth does not affect the operation principle of the proposed filter. However, as discussed before, the local defects (different rounding among teeth, lateral roughness, etc.) alter the phase of the light propagating through the grating, distorting the experimental filter response.

Subwavelength fiber-chip grating couplers were used to inject and extract the light from the chip with cleaved single mode (SMF-28) optical fibers \cite{halir2009waveguide, benedikovic2015subwavelength}. These grating couplers were optimized to minimize back-reflections for the TE polarized light, thereby minimizing the Fabry-Perot ripples in the measurements for proper analysis of transmission spectra of the filters.

\begin{figure}[htbp]
	\centering
	\centerline{\includegraphics*[width=8cm]{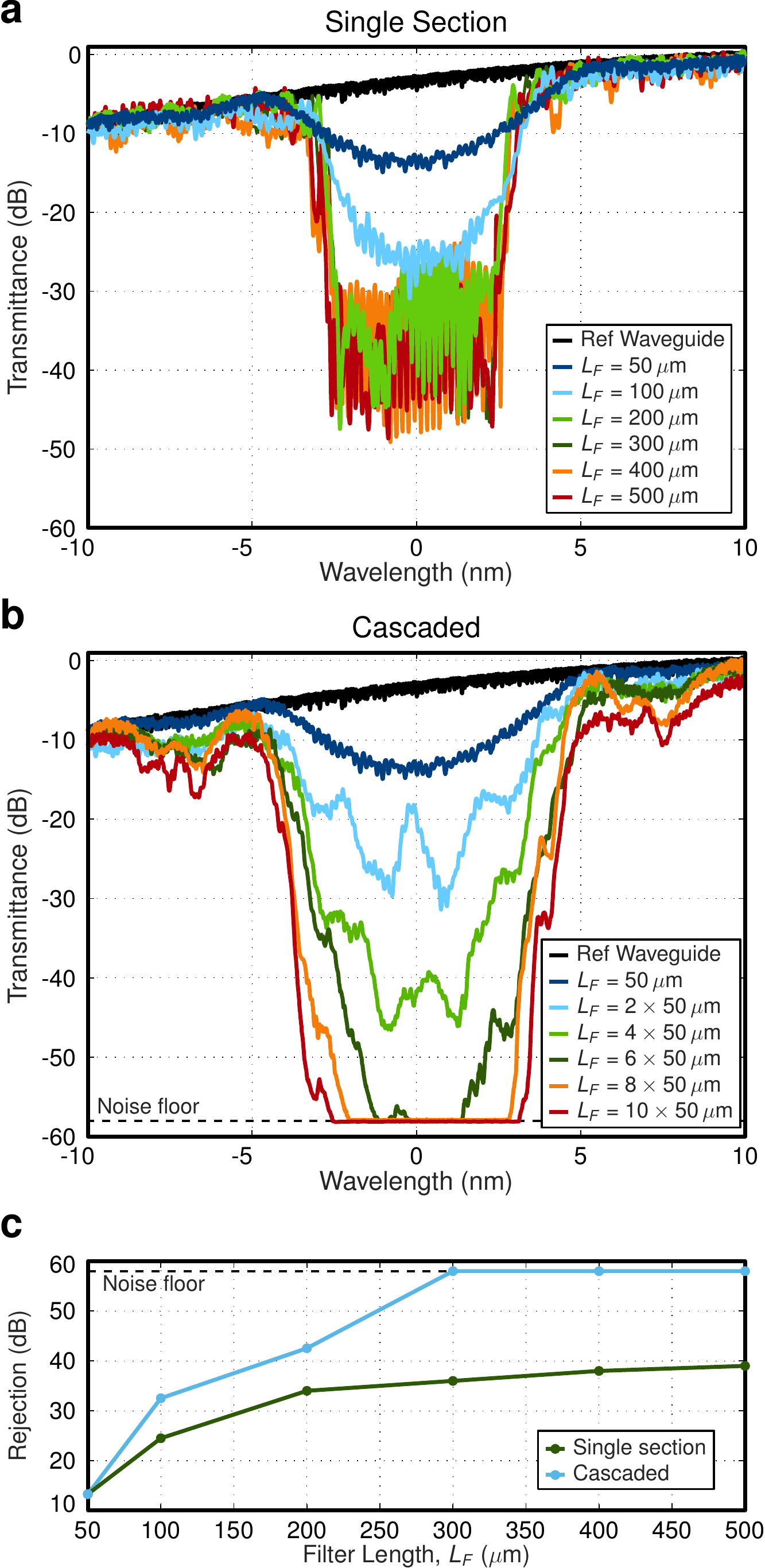}}
	\caption{Measured transmission spectra of (a) single section Bragg filters with increasing lengths, and (b) proposed cascaded Bragg filter with fixed section length, of 50\,$\mu$m, and number of sections ranging from 1 to 10. (c) Comparison of on-chip optical rejection as a function of filter length of single-section and proposed cascaded geometry, showing clear rejection saturation only for the single-section approach.}
	\label{fig:Saturation}
\end{figure}

First, a series of single-section shifted Bragg gratings with different lengths were considered to illustrate the effect of optical rejection saturation. As shown in Fig. \ref{fig:Saturation}(a), the rejection level saturates near 40 dB for filter lengths beyond 300\,$\mu$m. This comparatively weak rejection is mainly due to the strong index contrast between silicon and air that accentuates the detrimental effect of fabrication imperfections. We compared the transmission of the filters to that of a reference strip waveguide to demonstrate the low insertion loss of this kind of shifted geometry and show that the lower transmission at shorter wavelengths mainly arises from the response of the fiber-chip grating couplers. 

The potential of our new approach for non-coherent cascading is shown by the characterization of a set of cascaded shifted Bragg gratings separated by single-mode waveguides. The cascaded filters have the same total lengths as the single-section structures shown in Fig.  \ref{fig:Saturation}(a), but they are implemented by cascading multiple 50-$\mu$m-long grating sections. The 50\,$\mu$m section length has been chosen just as an illustrative example, being the main conclusions valid for other section lengths. The single-mode waveguides have a width of 400\,nm and a length of 20\,$\mu$m. We use 25-$\mu$m-long tapers to make adiabatic transition between Bragg gratings and input and output single-mode waveguides. As depicted in Fig. \ref{fig:Schema}(a), we included an S-bend (bending radius of 15\,$\mu$m) between each two grating sections to promote radiation of any remaining power carried by the back-reflected first order mode. The transmission spectra of the cascaded filters with total length ranging between 50\,$\mu$m and 500\,$\mu$m (comprising ten sections of 50\,$\mu$m length) are presented in Fig. \ref{fig:Saturation}(b). The proposed cascading approach yields a substantial increase in filter rejection, showing no clear evidence of saturation with the length. Note that for total lengths beyond 300\,$\mu$m the on-resonance transmission of the cascaded filter lies below the noise floor level of the automatic wavelength sweeping and detection system (CT400 from Yenista). 

In addition to a higher optical rejection, the cascaded filters exhibit a wider bandwidth. As discussed below, this wider rejection does not arise from relative wavelength shifts among different filter sections, but from the non-coherent nature of their interaction. From Eq. (\ref{eq:BW}) it follows that, for a given Bragg grating geometry (fixed $n_g$ and $\kappa$), filter bandwidth decreases with the length. This could be qualitatively explained from the point of view of Fourier transform, as a longer spatial perturbation results in a narrower spectral response. However, different sections in the modal-engineered filter do not interact coherently. Thus, the bandwidth of the proposed filter is not determined by the total length, but by the length of the sections. This non-coherent interaction effect can be observed in Fig. \ref{fig:Saturation}(b), showing that filters comprising 50-$\mu$m-long grating sections have a similar bandwidth, regardless the total length. 

\begin{figure}[htbp]
	\centering
	\centerline{\includegraphics*[width=8cm]{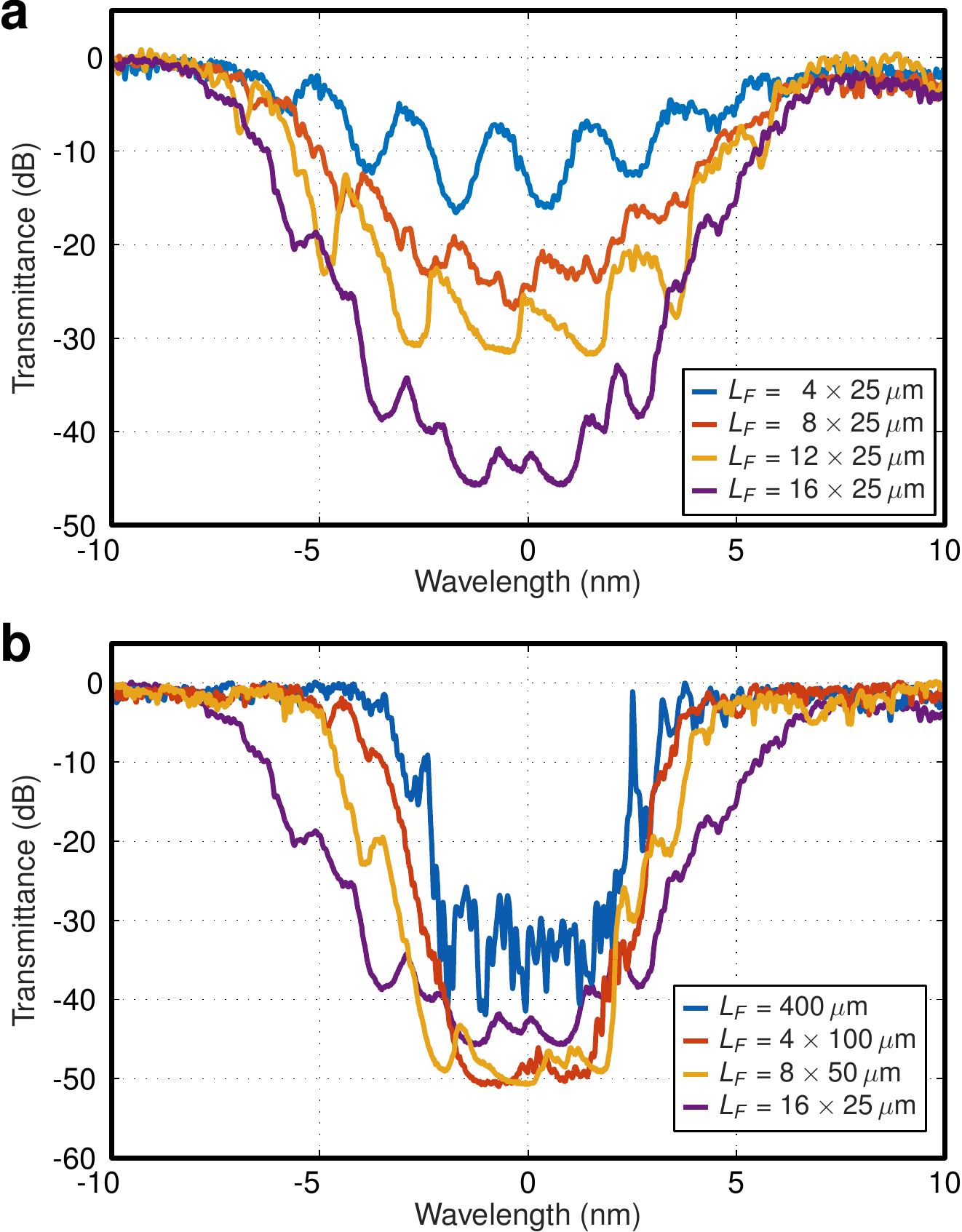}}
	\caption{Measured transmission spectra of proposed cascaded filter for (a) fixed section length, of 25\,$\mu$m, and increasing number of sections, showing no clear evidence of bandwidth increase with the number of sections, and (b) fixed total filter length, of 400\,$\mu$m, and decreasing section length. Filter bandwidth is determined by section length rather than by total length due to non-coherent cascading.}
	\label{fig:NonCoherent}
\end{figure}

Aiming at confirming the incoherent cascading in the proposed approach, we fabricated and characterized two different sets of filters. First, we fixed a section length of 25\,$\mu$m and cascaded different number of sections, from 4 to 16, resulting in total filter lengths between 100\,$\mu$m and 400\,$\mu$m. Measured spectra, shown in Fig. \ref{fig:NonCoherent}(a), demonstrate that the bandwidth of the cascaded filter does not significantly increase with an increasing number of sections. Hence, relative wavelength shifts can be discarded as the major reason for the wider bandwidth in the cascaded filters. This result is indeed expected, as all sections of the cascaded filter are fabricated under similar conditions, thus suffering from similar and random under/over etching effects. Then, any deviation of the central Bragg wavelength will be similar for all filter sections, yielding minimal relative shifts.

Then, we implemented the same total filter length, of 400\,$\mu$m, by a single-section device, and by cascading 4, 6 and 8 sections of 100\,$\mu$m, 50\,$\mu$m 25\,$\mu$m lengths, respectively. The different measured spectra are presented in Fig. \ref{fig:NonCoherent}(b). It can be noticed that, even if the total filter length is always the same, the filter bandwidth increases with decreasing section length. These results clearly demonstrate the incoherent cascading in the proposed geometry, as the total filter bandwidth is mainly determined by the bandwidth of the individual Bragg gratings sections, rather than by the total filter length. 

\begin{figure}[htbp]
	\centering
	\includegraphics[width=8cm]{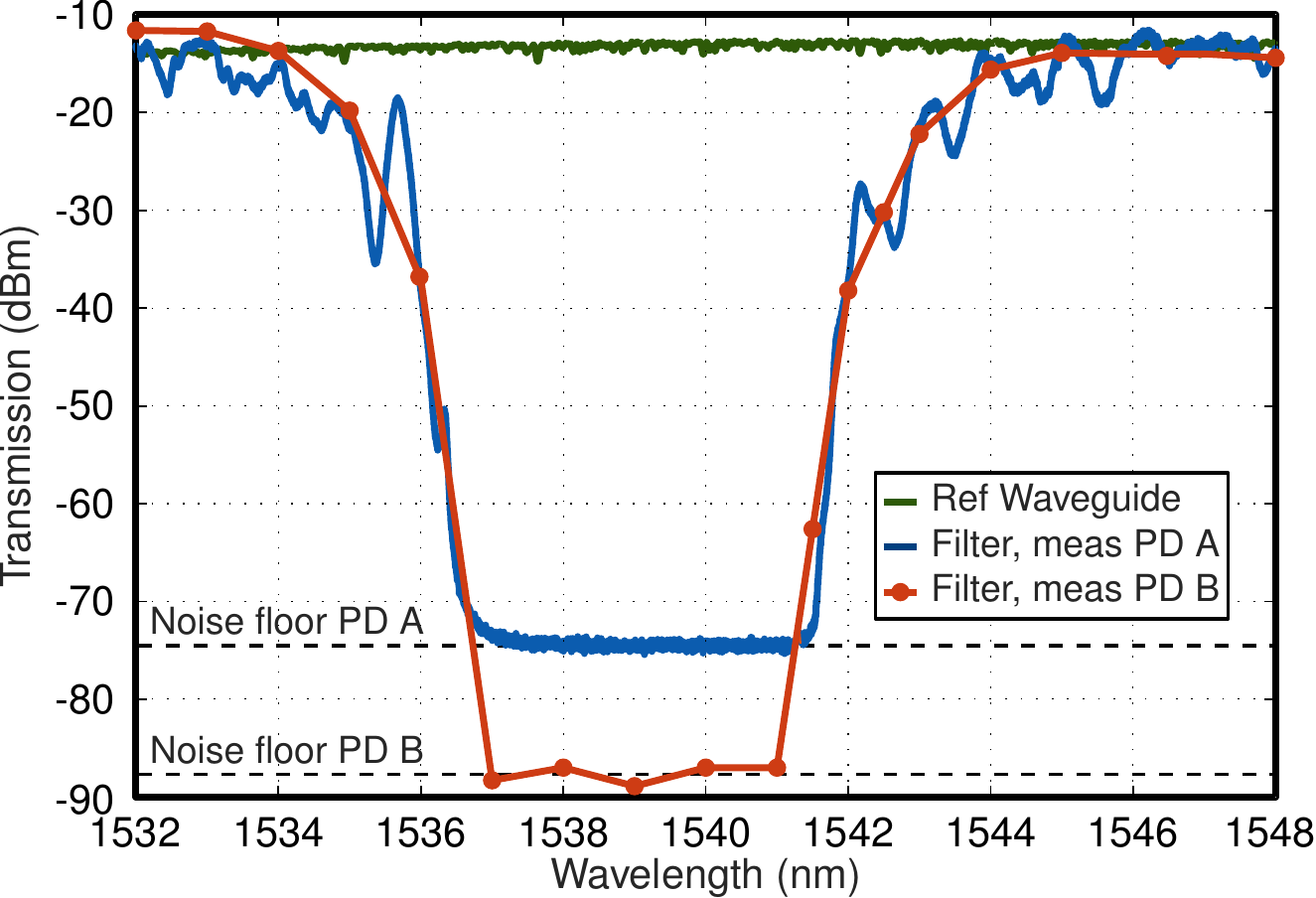}
	\caption{Transmission spectrum of cascaded shifted Bragg filter with total length of 2.5 mm, implemented by 10 modal-engineered Bragg grating sections of 250\,$\mu$m length. Measurements are performed with automatic wavelength sweep and detection system (CT400 form Yenista), PD A,  and high-sensitivity photo-detector, PD B, in OSA (Anritsu MS9710B). Transmission of 3.5-mm-long strip waveguide is shown for comparison.}
	\label{fig:deep}
\end{figure}

Finally, to demonstrate the remarkably large rejection capabilities of the proposed broken-coherency approach, we implemented a 2.5-mm-long filter, comprising 10 modal-engineered Bragg sections of 250\,$\mu$m length. The response of the filters was characterized using an automatic wavelength sweep and detection system (CT400 from Yenista) and the high-sensitivity photo-detector in  OSA Anritsu MS9710B (see Methods). For comparison, we included the response of a reference strip waveguide of 3.5 mm length. This length includes the grating lengths (total of 2.5 mm), and the lengths of the input/output tapers S-bends and single-mode sections. The proposed filter exhibits negligible off-band insertion loss, within the variability determined by fiber alignment precision and fabrication tolerances. The Bragg on-resonance transmission level of the cascaded filter lies within the noise floor of the OSA, with (at least) 80\,dB of on-chip optical rejection. This is, to the best of our knowledge, the largest rejection experimentally demonstrated for a silicon waveguide Bragg filter. We used a fiber circulator at the input of the filter to collect all back-reflections. We retrieved a broadband and quasi-flat signal with no signature of the Bragg resonance and nearly -40\,dBm level (mainly arising from reflections in gratings and backscattering in waveguide roughness). This result further confirms that Bragg back-reflections are effectively radiated away in the single-mode waveguide sections.

\section*{Discussion}

In conclusion, we have proposed and experimentally demonstrated a new technology-independent strategy to preclude coherent interaction between cascaded Bragg filters. This strategy provides a dramatic optical rejection increase, overcoming on of the major on-chip performance limitations.  While maximum rejection level in conventional wavelength filters is seriously limited by phase errors arising from fabrication imperfections, our approach allows effective cascading of low rejection level stages without the need for any active tuning. The innovative concept is to separate multi-mode Bragg grating sections by single-mode waveguides. We engineer the grating to yield Bragg back-reflections propagating in a high-order spatial mode, which is radiated away in the single-mode waveguides. This way, different filter sections are completely independent with no phase relationship, allowing effective optical rejection accumulation, even in the presence of phase errors. Based on this concept, we have experimentally demonstrated on-chip non-coherent cascading of multi-stage silicon Bragg filters, and have implemented a notch wavelength filter with an optical rejection higher than 80\,dB. This is the largest optical rejection ever reported for an all-passive silicon photonic wavelength filter. The approach proposed here can be translated to any other integrated photonic technology, as long as it allows the realization of multi-mode Bragg gratings and single-mode waveguides. Furthermore, the broken-coherency strategy releases new degrees of freedom to tailor the shape of on-chip wavelength filters. More specifically, the bandwidth of the cascaded Bragg filter is mainly determined by the length of the single section, while the rejection depth is set by the number of sections. This unique capability to overcome bandwidth-rejection trade-off in conventional Bragg filters opens exciting opportunities for the development of efficient and fabrication tolerant Si wavelength filters, with a great potential for integrated nonlinear applications, e.g. next generation Si-based photon-pair sources for quantum photonic circuits.

\section*{Methods}

\subsection*{Device fabrication and experimental characterization}
We fabricated the different sets of shifted Bragg filters in SOI wafers comprising a $220$\,nm thick silicon and a $ 2\,\mu$m buried oxide layer, using an electron beam lithography (Nanobeam NB-4 system 80kV) with a step size of $5$\,nm. Dry and inductively coupled plasma etching ($\mathrm{SF_6}/\mathrm{C_4F_8}$) were used to define the patterns. We used a tunable laser source from Yenista, providing 10\,dBm output power around 1550\,nm wavelength. To collect the filter spectra, we used automatic high-resolution wavelength scan with Yenista CT400 (noise floor of about $ -75\,\mathrm{dBm}$) and point-by-point scan with optical spectrum analyzer (OSA Anritsu MS9710B, with noise floor around $-90\,\mathrm{dBm}$). 


\begin{thebibliography}{10}
	\expandafter\ifx\csname url\endcsname\relax
	\def\url#1{\texttt{#1}}\fi
	\expandafter\ifx\csname urlprefix\endcsname\relax\def\urlprefix{URL }\fi
	\expandafter\ifx\csname doiprefix\endcsname\relax\def\doiprefix{DOI }\fi
	\providecommand{\bibinfo}[2]{#2}
	\providecommand{\eprint}[2][]{\url{#2}}
	
	\bibitem{boeuf2016silicon}
	\bibinfo{author}{Boeuf, F.} \emph{et~al.}
	\newblock \bibinfo{journal}{\bibinfo{title}{Silicon photonics R\&D and
			manufacturing on 300-mm wafer platform}}.
	\newblock {\emph{\JournalTitle{J. Light. Technol.}}}
	\textbf{\bibinfo{volume}{34}}, \bibinfo{pages}{286--295}
	(\bibinfo{year}{2016}).
	
	\bibitem{fernandez2016last}
	\bibinfo{author}{Fern{\'a}ndez~Gavela, A.},
	\bibinfo{author}{Grajales~Garc{\'\i}a, D.}, \bibinfo{author}{Ramirez, J.~C.}
	\& \bibinfo{author}{Lechuga, L.~M.}
	\newblock \bibinfo{journal}{\bibinfo{title}{Last advances in silicon-based
			optical biosensors}}.
	\newblock {\emph{\JournalTitle{Sensors}}} \textbf{\bibinfo{volume}{16}},
	\bibinfo{pages}{285} (\bibinfo{year}{2016}).
	
	\bibitem{janz2013photonic}
	\bibinfo{author}{Janz, S.} \emph{et~al.}
	\newblock \bibinfo{journal}{\bibinfo{title}{Photonic wire biosensor microarray
			chip and instrumentation with application to serotyping of Escherichia
			coliisolates}}.
	\newblock {\emph{\JournalTitle{Opt. Express}}} \textbf{\bibinfo{volume}{21}},
	\bibinfo{pages}{4623--4637} (\bibinfo{year}{2013}).
	
	\bibitem{knill2001scheme}
	\bibinfo{author}{Knill, E.}, \bibinfo{author}{Laflamme, R.} \&
	\bibinfo{author}{Milburn, G.~J.}
	\newblock \bibinfo{journal}{\bibinfo{title}{A scheme for efficient quantum
			computation with linear optics}}.
	\newblock {\emph{\JournalTitle{Nature}}} \textbf{\bibinfo{volume}{409}},
	\bibinfo{pages}{46} (\bibinfo{year}{2001}).
	
	\bibitem{mazeas2016high}
	\bibinfo{author}{Mazeas, F.} \emph{et~al.}
	\newblock \bibinfo{journal}{\bibinfo{title}{High-quality photonic entanglement
			for wavelength-multiplexed quantum communication based on a silicon chip}}.
	\newblock {\emph{\JournalTitle{Opt. Express}}} \textbf{\bibinfo{volume}{24}},
	\bibinfo{pages}{28731--28738} (\bibinfo{year}{2016}).
	
	\bibitem{cheben2015broadband}
	\bibinfo{author}{Cheben, P.} \emph{et~al.}
	\newblock \bibinfo{journal}{\bibinfo{title}{Broadband polarization independent
			nanophotonic coupler for silicon waveguides with ultra-high efficiency}}.
	\newblock {\emph{\JournalTitle{Opt. Express}}} \textbf{\bibinfo{volume}{23}},
	\bibinfo{pages}{22553--22563} (\bibinfo{year}{2015}).
	
	\bibitem{ziebell201240}
	\bibinfo{author}{Ziebell, M.} \emph{et~al.}
	\newblock \bibinfo{journal}{\bibinfo{title}{40 Gbit/s low-loss silicon
			optical modulator based on a PIPIN diode}}.
	\newblock {\emph{\JournalTitle{Opt. Express}}} \textbf{\bibinfo{volume}{20}},
	\bibinfo{pages}{10591--10596} (\bibinfo{year}{2012}).
	
	\bibitem{vivien2012zero}
	\bibinfo{author}{Vivien, L.} \emph{et~al.}
	\newblock \bibinfo{journal}{\bibinfo{title}{Zero-bias 40 Gbit/s germanium
			waveguide photodetector on silicon}}.
	\newblock {\emph{\JournalTitle{Opt. Express}}} \textbf{\bibinfo{volume}{20}},
	\bibinfo{pages}{1096--1101} (\bibinfo{year}{2012}).
	
	\bibitem{jouguet2013experimental}
	\bibinfo{author}{Jouguet, P.}, \bibinfo{author}{Kunz-Jacques, S.},
	\bibinfo{author}{Leverrier, A.}, \bibinfo{author}{Grangier, P.} \&
	\bibinfo{author}{Diamanti, E.}
	\newblock \bibinfo{journal}{\bibinfo{title}{Experimental demonstration of
			long-distance continuous-variable quantum key distribution}}.
	\newblock {\emph{\JournalTitle{Nature Photon.}}}
	\textbf{\bibinfo{volume}{7}}, \bibinfo{pages}{378--381}
	(\bibinfo{year}{2013}).
	
	\bibitem{azzini2012ultra}
	\bibinfo{author}{Azzini, S.} \emph{et~al.}
	\newblock \bibinfo{journal}{\bibinfo{title}{Ultra-low power generation of twin
			photons in a compact silicon ring resonator}}.
	\newblock {\emph{\JournalTitle{Opt. Express}}} \textbf{\bibinfo{volume}{20}},
	\bibinfo{pages}{23100--23107} (\bibinfo{year}{2012}).
	
	\bibitem{jiang2015silicon}
	\bibinfo{author}{Jiang, W.~C.}, \bibinfo{author}{Lu, X.},
	\bibinfo{author}{Zhang, J.}, \bibinfo{author}{Painter, O.} \&
	\bibinfo{author}{Lin, Q.}
	\newblock \bibinfo{journal}{\bibinfo{title}{Silicon-chip source of bright
			photon pairs}}.
	\newblock {\emph{\JournalTitle{Opt. Express}}} \textbf{\bibinfo{volume}{23}},
	\bibinfo{pages}{20884--20904} (\bibinfo{year}{2015}).
	
	\bibitem{grassani2015micrometer}
	\bibinfo{author}{Grassani, D.} \emph{et~al.}
	\newblock \bibinfo{journal}{\bibinfo{title}{Micrometer-scale integrated silicon
			source of time-energy entangled photons}}.
	\newblock {\emph{\JournalTitle{Optica}}} \textbf{\bibinfo{volume}{2}},
	\bibinfo{pages}{88--94} (\bibinfo{year}{2015}).
	
	\bibitem{piekarek2017high}
	\bibinfo{author}{Piekarek, M.} \emph{et~al.}
	\newblock \bibinfo{journal}{\bibinfo{title}{High-extinction ratio integrated
			photonic filters for silicon quantum photonics}}.
	\newblock {\emph{\JournalTitle{Opt. Lett.}}} \textbf{\bibinfo{volume}{42}},
	\bibinfo{pages}{815--818} (\bibinfo{year}{2017}).
	
	\bibitem{wang2011uniform}
	\bibinfo{author}{Wang, X.}, \bibinfo{author}{Shi, W.}, \bibinfo{author}{Vafaei,
		R.}, \bibinfo{author}{Jaeger, N.~A.} \& \bibinfo{author}{Chrostowski, L.}
	\newblock \bibinfo{journal}{\bibinfo{title}{Uniform and sampled Bragg gratings
			in SOI strip waveguides with sidewall corrugations}}.
	\newblock {\emph{\JournalTitle{IEEE Photon. Technol. Lett.}}}
	\textbf{\bibinfo{volume}{23}}, \bibinfo{pages}{290--292}
	(\bibinfo{year}{2011}).
	
	\bibitem{perez2017optical}
	\bibinfo{author}{P{\'e}rez-Galacho, D.} \emph{et~al.}
	\newblock \bibinfo{journal}{\bibinfo{title}{Optical pump-rejection filter based
			on silicon sub-wavelength engineered photonic structures}}.
	\newblock {\emph{\JournalTitle{Opt. Lett.}}} \textbf{\bibinfo{volume}{42}},
	\bibinfo{pages}{1468--1471} (\bibinfo{year}{2017}).
	
	\bibitem{klitis2017high}
	\bibinfo{author}{Klitis, C.}, \bibinfo{author}{Cantarella, G.},
	\bibinfo{author}{Strain, M.~J.} \& \bibinfo{author}{Sorel, M.}
	\newblock \bibinfo{journal}{\bibinfo{title}{High-extinction-ratio TE/TM
			selective Bragg grating filters on silicon-on-insulator}}.
	\newblock {\emph{\JournalTitle{Opt. Lett.}}} \textbf{\bibinfo{volume}{42}},
	\bibinfo{pages}{3040--3043} (\bibinfo{year}{2017}).
	
	\bibitem{xia2007ultra}
	\bibinfo{author}{Xia, F.}, \bibinfo{author}{Rooks, M.},
	\bibinfo{author}{Sekaric, L.} \& \bibinfo{author}{Vlasov, Y.}
	\newblock \bibinfo{journal}{\bibinfo{title}{Ultra-compact high order ring
			resonator filters using submicron silicon photonic wires for on-chip optical
			interconnects}}.
	\newblock {\emph{\JournalTitle{Opt. Express}}} \textbf{\bibinfo{volume}{15}},
	\bibinfo{pages}{11934--11941} (\bibinfo{year}{2007}).
	
	\bibitem{dong2010ghz}
	\bibinfo{author}{Dong, P.} \emph{et~al.}
	\newblock \bibinfo{journal}{\bibinfo{title}{GHz-bandwidth optical filters based
			on high-order silicon ring resonators}}.
	\newblock {\emph{\JournalTitle{Opt. Express}}} \textbf{\bibinfo{volume}{18}},
	\bibinfo{pages}{23784--23789} (\bibinfo{year}{2010}).
	
	\bibitem{ding2011bandwidth}
	\bibinfo{author}{Ding, Y.} \emph{et~al.}
	\newblock \bibinfo{journal}{\bibinfo{title}{Bandwidth and wavelength-tunable
			optical bandpass filter based on silicon microring-MZI structure}}.
	\newblock {\emph{\JournalTitle{Opt. Express}}} \textbf{\bibinfo{volume}{19}},
	\bibinfo{pages}{6462--6470} (\bibinfo{year}{2011}).
	
	\bibitem{horst2013cascaded}
	\bibinfo{author}{Horst, F.} \emph{et~al.}
	\newblock \bibinfo{journal}{\bibinfo{title}{Cascaded Mach-Zehnder wavelength
			filters in silicon photonics for low loss and flat pass-band WDM (de-)
			multiplexing}}.
	\newblock {\emph{\JournalTitle{Opt. Express}}} \textbf{\bibinfo{volume}{21}},
	\bibinfo{pages}{11652--11658} (\bibinfo{year}{2013}).
	
	\bibitem{wang2015subwavelength}
	\bibinfo{author}{Wang, J.}, \bibinfo{author}{Glesk, I.} \&
	\bibinfo{author}{Chen, L.}
	\newblock \bibinfo{journal}{\bibinfo{title}{Subwavelength grating Bragg grating
			filters in silicon-on-insulator}}.
	\newblock {\emph{\JournalTitle{Electron. Lett.}}}
	\textbf{\bibinfo{volume}{51}}, \bibinfo{pages}{712--714}
	(\bibinfo{year}{2015}).
	
	\bibitem{zou2016tunable}
	\bibinfo{author}{Zou, Z.}, \bibinfo{author}{Zhou, L.}, \bibinfo{author}{Wang,
		M.}, \bibinfo{author}{Wu, K.} \& \bibinfo{author}{Chen, J.}
	\newblock \bibinfo{journal}{\bibinfo{title}{Tunable spiral Bragg gratings in
			60-nm-thick silicon-on-insulator strip waveguides}}.
	\newblock {\emph{\JournalTitle{Opt. Express}}} \textbf{\bibinfo{volume}{24}},
	\bibinfo{pages}{12831--12839} (\bibinfo{year}{2016}).
	
	\bibitem{lu2017performance}
	\bibinfo{author}{Lu, Z.} \emph{et~al.}
	\newblock \bibinfo{journal}{\bibinfo{title}{Performance prediction for silicon
			photonics integrated circuits with layout-dependent correlated manufacturing
			variability}}.
	\newblock {\emph{\JournalTitle{Opt. Express}}} \textbf{\bibinfo{volume}{25}},
	\bibinfo{pages}{9712--9733} (\bibinfo{year}{2017}).
	
	\bibitem{ong2013ultra}
	\bibinfo{author}{Ong, J.~R.}, \bibinfo{author}{Kumar, R.} \&
	\bibinfo{author}{Mookherjea, S.}
	\newblock \bibinfo{journal}{\bibinfo{title}{Ultra-high-contrast and
			tunable-bandwidth filter using cascaded high-order silicon microring
			filters}}.
	\newblock {\emph{\JournalTitle{IEEE Photon. Technol. Lett.}}}
	\textbf{\bibinfo{volume}{25}}, \bibinfo{pages}{1543--1546}
	(\bibinfo{year}{2013}).
	
	\bibitem{liu2009photonic}
	\bibinfo{author}{Liu, J.-M.}
	\newblock \emph{\bibinfo{title}{Photonic devices}}
	(\bibinfo{publisher}{Cambridge University Press}, \bibinfo{year}{2009}).
	
	
	
	
	\bibitem{CMT}
	\bibinfo{author}{Yariv, A.}
	\newblock \bibinfo{journal}{\bibinfo{title}{Coupled-mode theory for guided-wave
			optics}}.
	\newblock {\emph{\JournalTitle{IEEE J. Quantum Electron.}}}
	\textbf{\bibinfo{volume}{9}}, \bibinfo{pages}{919--933}
	(\bibinfo{year}{1973}).
	
	\bibitem{Shifted}
	\bibinfo{author}{Wang, X.} \emph{et~al.}
	\newblock \bibinfo{journal}{\bibinfo{title}{Precise control of the coupling
			coefficient through destructive interference in silicon waveguide Bragg
			gratings}}.
	\newblock {\emph{\JournalTitle{Opt. Lett.}}} \textbf{\bibinfo{volume}{39}},
	\bibinfo{pages}{5519--5522} (\bibinfo{year}{2014}).
	
	\bibitem{qiu2016silicon}
	\bibinfo{author}{Qiu, H.} \emph{et~al.}
	\newblock \bibinfo{journal}{\bibinfo{title}{Silicon band-rejection and
			band-pass filter based on asymmetric Bragg sidewall gratings in a multimode
			waveguide}}.
	\newblock {\emph{\JournalTitle{Opt. Lett.}}} \textbf{\bibinfo{volume}{41}},
	\bibinfo{pages}{2450--2453} (\bibinfo{year}{2016}).
	
	\bibitem{Lumerical}
	\bibinfo{title}{FDTD solutions, Lumerical solutions, inc.,
		http://www.lumerical.com}.
	
	\bibitem{halir2009waveguide}
	\bibinfo{author}{Halir, R.} \emph{et~al.}
	\newblock \bibinfo{journal}{\bibinfo{title}{Waveguide grating coupler with
			subwavelength microstructures}}.
	\newblock {\emph{\JournalTitle{Opt. Lett.}}} \textbf{\bibinfo{volume}{34}},
	\bibinfo{pages}{1408--1410} (\bibinfo{year}{2009}).
	
	\bibitem{benedikovic2015subwavelength}
	\bibinfo{author}{Benedikovic, D.} \emph{et~al.}
	\newblock \bibinfo{journal}{\bibinfo{title}{Subwavelength index engineered
			surface grating coupler with sub-decibel efficiency for 220-nm
			silicon-on-insulator waveguides}}.
	\newblock {\emph{\JournalTitle{Opt. Express}}} \textbf{\bibinfo{volume}{23}},
	\bibinfo{pages}{22628--22635} (\bibinfo{year}{2015}).
	
\end{thebibliography}

\section*{Acknowledgements (not compulsory)}

This work has been partially funded by the Agence Nationale de la Recherche (ANR-SITQOM-15-CE24-0005) and the H2020 European Research Council (ERC) (ERC POPSTAR 647342). The fabrication of the device was performed at the Plateforme de Micro-Nano-Technologie/C2N, which is partially funded by the Conseil G\'{e}n\'{e}ral de l’Essonne. This work was partly supported by the French RENATECH network.

$^{\bigtriangleup}$ D. P\'{e}rez Galacho is currently at Photonics Reseach Labs, iTEAM Research Institute, Universitat Politecnica de Valencia, Spain.

\section*{Author contributions statement}

C.A.R. proposed the concept. D.O., D.P.G. and E.C. designed the devices. X.L.R. and D.O. fabricated the samples. D.O., F.M., C.A.R., O.A. and L.L. performed the experimental characterization. D.O., F.M., S.T., L.L., D. M.-M., L.V., E.C. discussed the results and wrote the manuscript.

\section*{Additional information}

Competing Interests: The authors declare that they have no competing interests.

\end{document}